# Mechanism of Instrumental Game Theory in The Legal Process *via* Stochastic Options Pricing Induction

*Kwadwo Osei Bonsu*

*(corresponding author)*

[k.oseibonsu@pop.zjgsu.edu.cn,](mailto:k.oseibonsu@pop.zjgsu.edu.cn)

*School of Economics,*
*School of Law and Intellectual Property*
*Zhejiang Gongshang University*

*Professor Shoucan Chen*
*School of Law and Intellectual Property*
*Zhejiang Gongshang University*

*Abstract*

Economic theory has provided an estimable intuition in understanding the perplexing ideologies in law, in the areas of economic law, tort law, contract law, procedural law and many others. Most legal systems require the parties involved in a legal dispute to exchange information through a process called discovery. The purpose is to reduce the relative optimisms developed by asymmetric information between the parties. Like a head or tail phenomenon in stochastic processes, uncertainty in the adjudication affects the decisions of the parties in a legal negotiation. This paper therefore applies the principles of aleatory analysis to determine how negotiations fail in the legal process, introduce the axiological concept of optimal transaction cost and formulates a numerical methodology based on backwards induction and stochastic options pricing economics in estimating the reasonable and fair bargain in order to induce settlements thereby increasing efficiency and reducing social costs.

**Formulation of Hypothesis and Research Design**

From Adam Smith's comments on the economic influence on mercantilist legislation to the 20th Century R. Coase, G. Calabresi, through to the Chicago school and the contemporary philosophy of the economic interpretation of various aspects of the legal scholarship, we've seen how the mass elixir of economic and mathematical intuition has shed light on the legal philosophy and process both in academia and legal practice.

Richard A. Posner shows the difference between utilitarianism and economics as many legal scholars at the time criticized the utilitarian philosophy of Economic Analysis of Law [1]. He also shows that the economic norm of wealth maximization provides a substantial basis for a normative theory of law as opposed to shear utilitarianism, hence the system of wealth maximization transfers wealth to those who have productive assets in the form of time, goods, or skill. Those who have little or no productive assets have little or no claim on the assets of others hence little or no wealth is transferred to them[2]. Posner's assertions provide a basis for understanding the value of exchange not only in the law but also in trade, that one needs to provide something in order to gain another in return.

In a legal dispute, parties engage in a series of negotiations so as to resolve the issue without trial. The parties need to present a fair and reasonable bargain in order to induce the WTA (willingness to accept) of the plaintiff and the WTP(willingness to pay) of the defendant. Cooperation can only be attained when the WTA of the plaintiff is less than or equal to the WTP of the defendant.

Professor Ronald H. Coase in his famous Coase Theorem posits that if there are no transaction costs parties in a dispute will reach an efficient outcome in the presence of externalities regardless of how property rights are allocated[3]. The law should be structured to minimize the impediments that hinder cooperation and when cooperation fails the the law should aim to minimize the harm caused by failures in private negotiations. Therefore the Normative Coase-Hobbes Theorem suggests that in order to reduce the harm caused by the failures in private agreements and to render transaction costs irrelevant , the law should allocate the right to the party which values it the most[4]. Hobbes' original philosophy is presented in his work , Leviathan in 1651[5].

Coming up with coherent models in transaction cost analysis has been very difficult, this paper however seeks to assert an optimal transaction cost theory that does not only shed light on Coase Theorem but also proposes a more applicable analysis of the transaction cost in the legal process [6].

Arnold likened the operation of the judicial system to a theatrical performance where the judges are actors from whose play lines moral lessons could be learned , the litigants who have sponsored the show are not welcome to the show because the principles of substantive law is constantly aimed at preventing or reducing the rate of litigation by referring the litigants to renegotiations ,

---

[1] Even today many legal scholars still misconstrue the concepts of utilitarianism and wealth maximization in the analysis of law. See; Richard A. Posner, "Utilitarianism, Economics, and Legal Theory," The Journal of Legal Studies 8, no. 1 (Jan., 1979): 103-140

[2] Richard A. Posner , "The Ethical And Political Basis of The Efficiency Norm in Common Law Adjudication," 8 Hofstra L. Rev. 487 1979-1980

[3] Ronald H. Coase , The Problem of Social Cost, 3 J. Law and Econ. 1 (1960)

[4] Cooter, The Cost of Coase, 11 J. Legal Stud. 1 (1982)

[5] The original work by Thomas Hobbes in 1651 is regarded as one of the earliest and most influential examples of social contract theory , see;
Hobbes, T. (1996). Hobbes: Leviathan: Revised student edition(Cambridge Texts in the History of Political Thought) (R.Tuck,Ed.).Cambridge: Cambridge University Press. doi:10.1017/CBO9780511808166

[6] Steven Tadelis and Oliver Williamson, "Transaction Cost Economics," University of California , Berkeley, November 14 , 2010 ,available online at;
https://pdfs.semanticscholar.org/e4e8/a0486808360d056dbe212f7424273558538c.pdf

arbitration or telling them to resort to the action of a commission[7]. The question Arnold was putting across was that if the play is not sponsored by the litigants then the audience will loose the opportunity to learn the moral lessons of the play. Meaning, the occurrence of trials produce precedent for resolving future cases, so if the law is structured so as to reduce the number of trials by inducing private agreements then less precedent might be created.

Less precedent also means more uncertainty in the future judgments which will further induce more private cooperation especially with risk averse litigants. Nevertheless, when precedent is lacking, courts find it difficult to pass judgments that reflect with the social norms so injurers exercise less precaution thereby resulting in more trials. The focus of this paper however is not to propose a solution to this dilemma but to show how the uncertainty of judgment can influence the choices made by litigants in private agreements.

There are three sections, Section 1 formulates a numerical instrumentation in analyzing the reasonable bargain under symmetric and asymmetric information, proposes the optimal transaction cost theory and discusses the effect of time on private agreements in the legal process. Section 2 devises a quantitative methodology in analyzing the factors that determine the axiological threshold in the bargain position of litigants and uses a probabilistic approach in estimating the 'fair bargain' under uncertainty of the outcome of judgment; a method formulated from stochastic options pricing economics.

The last section, Section 3 then tests the fair bargain formularization through conceptual analysis of the legal process to highlight the application of the methods developed in this paper.

## Section 1: Game Theoretic Analysis of the Legal process

There are conflicts between Posner's "value of exchange" and Coase-Hobbes' "allocation of rights to the party which values the most" , this is why we need a theory to draw us from the allocative paradigms to a more self-reliant approach.

Litigants in a dispute as rational beings seek to maximize their benefits and reduce their losses as much as possible. This utility maximization (though not exactly utilitarianism) leads them to make decisions based on strategies in order to counter their opponent just as any other game within the purview of game theory.

Prisoners' dilemma is a famous game in which players are not allowed to communicate with each other, a perfect example of a game under asymmetric information. If the players could negotiate they would come up with a strategy that would maximize each players benefit. Fortunately, the legal process is just the opposite, litigants are taken through the discovery process and are encouraged to share information. Albeit the presence of discovery we still see trials taking place after negotiations, some cases just a few , others several negotiations still do not result in settlements.

Information alone is not enough to induce private agreements, players in the game need a reasonable and fair bargain from the opponent to result in cooperation. In a 1994 lecture, Randal C.

---

[7] Arnold, Thurman W., "The Role of Substantive Law and Procedure in the Legal Process" (1932). Faculty Scholarship Series. Paper 4258, available online at;
http://digitalcommons.law.yale.edu/fss_papers/4258

Parker explained that even though players in a game could have dominant strategies a game could end up in a single or multiple Nash equilibria ; as a consequence, players would then use pure strategy, that is none of the players would be playing in a probabilistic fashion[8]. Under symmetric information litigants may devise other ways of playing the game even though they will be using pure strategies.

See figure (1) below:

---

[8] For an extensive inquiry into how players in a game use pure strategies in equilibrium and the relationship between such games and the legal process see; Randal C. Picker, "An Introduction to Game Theory and the Law" (Coase-Sandor Institute for Law & Economics Working Paper No. 22, 1994)

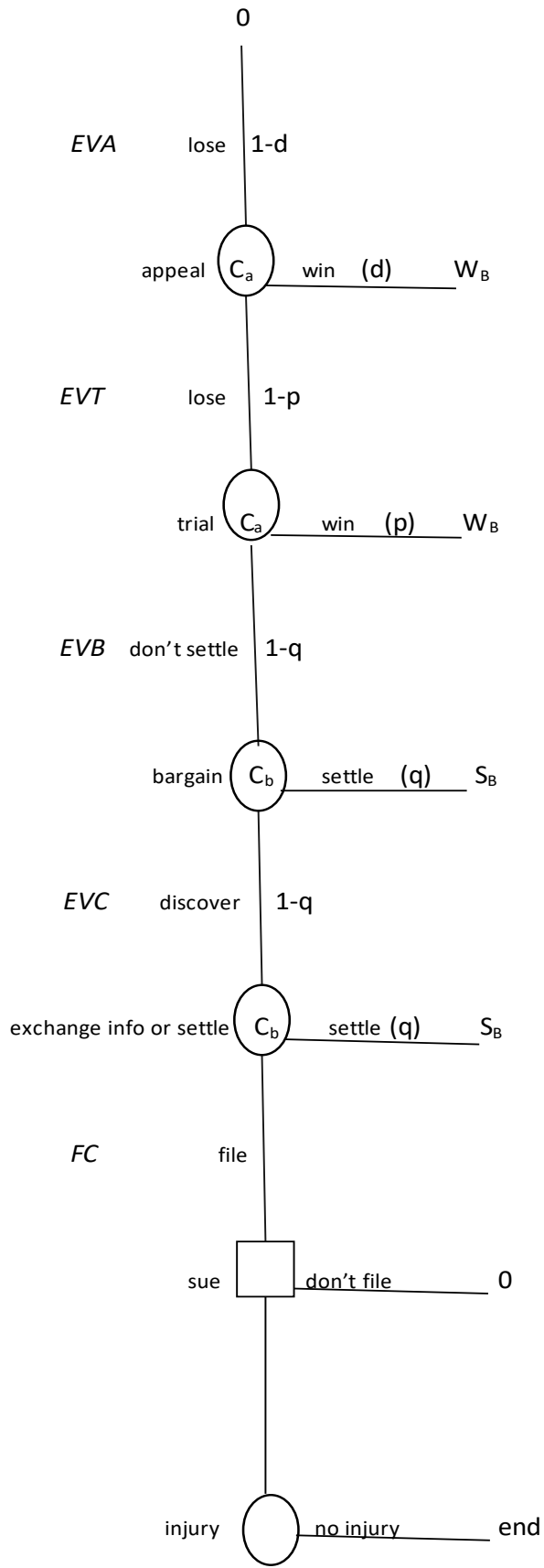

Figure (1)

Figure (1) represents the decisions and outcomes in a legal process; the defendant either causes injury to the plaintiff or no injury. If no injury game ends (injury here represents all kinds of harm including breach of contract ,tort and many others) . If there is an injury the plaintiff decides to file a suit or let it go for a payoff of zero. A suit results in either exchange of information through discovery process or outright settlement. There is a probability q for settlement with a payoff $S_B$ (let's call it benefit of settlement), and a probability of 1-q when discovery results in a non-settlement, meaning negotiation is needed so the next stage is a bargain. From the bargain, there is still a probability q for the bargain to achieve settlement and 1-q when negotiation fails leading to a trial. The plaintiff has a probability p of winning and 1-p for losing the trial. $W_B$ is the payoff from winning at the trial, thus the judgment in favor of the plaintiff ( let's call it winning benefit). If the plaintiff wins at trail, the defendant may appeal, likewise if the plaintiff loses she may go for an appeal that results in either win with a probability d and payoff $W_B$, or lose with a probability 1-d and payoff zero. Either way the game ends because an appeal is to rectify the problems of the application of the law without further evidence or arguments being made. Also the outcome of the appeal isn't dependent probabilistically on the strategies of the litigants. $C_a$ and $C_b$ are the administration costs (including court fees, trials, and any costs incurred during the trial process) and bargain costs (including negotiations, and all costs incurred during discovery and bargaining process) respectively. For simplicity of the model, we assume all administration costs are the same for all trials and all costs of bargain are the same for all negotiations and exchange of information. Moreover, this paper uses the American rule where each party bears their own litigation costs regardless of whether they win or lose.

Expected values of each stage in the litigation process is deduced by backwards induction as follows;

EVA , expected value of the appeal ;
$$\rightarrow \quad EVA = (1-d) \times 0 + d(W_B) - C_a \quad\quad\quad (1)$$

EVT, expected value of trial ;
$$\rightarrow \quad EVT = (1-p) \times 0 + p(W_B) - C_a \quad\quad\quad (2)$$

EVB, expected value of bargain ;
$$\rightarrow \quad EVB = (1-q) \times (EVT) + q(S_B) - C_b \quad\quad\quad (3)$$

EVC, expected value of claim ;
$$\rightarrow \quad EVC = (1-q) \times (EVB) + q(S_B) - C_b \quad\quad\quad (4)$$

substituting equation (2) into (3) , we have ;

$$EVB = (1-q)[\, pW_B - C_a] + qS_B - C_b \quad\quad\quad (5)$$

substituting equation (5) into (4) , we have ;

$$EVC = (1-q)[ (1-q)[ pW_B - C_a] + qS_B - C_b] + qS_B - C_b$$

$\Rightarrow$

$$EVC = pW_B - pqW_B - C_a + qC_a + qS_B - C_b$$
$$- pqW_B + pq^2W_B + qC_a - q^2C_a - q^2S_B$$
$$+ qC_b + qS_B - C_b$$

$$EVC = W_B[p - 2pq + pq^2] + C_a[-1 + 2q - q^2]$$
$$+ S_B[2q - q^2] + C_b[-2 + q] \quad (6)$$

The threat value in a game is the payoff to a player in noncooperation. Legal scholars prefer the term 'go-it-alone value' due to the pugnacious nature of the word 'threat' especially when used in courts. In a legal bargain, the threat value, let's denote as TP is at least as much as the expected value of the claim which is at least equal to the cost of filing the claim ,FC.

Thus;

TP ≥ EVC ≥ FC

According to economic efficiency a rational litigant will only file a legal claim when the cost of filing a claim is at least equal to the expected value of claim.

Hence we can deduce that for a minimum threat value,

TP=EVC ;

$$TP = W_B\left[ p - 2pq + pq^2 \right] + C_a\left[ -1 + 2q - q^2 \right]$$
$$+ S_B\left[ 2q - q^2 \right] + C_b[-2 + q]$$

therefore,

$$TP = W_B[p(1-q)^2] - C_a(1-q)^2$$
$$+ S_B[q(2-q)] - C_b(2-q) \quad (7)$$

**Section 1.1 : Reasonable Bargain in a Legal Negotiation**

Let's take $B_N$ as the noncooperative bargain, thus threat value under no chance of settlement, q = 0 , from equation (7) we have ;

$$B_N = pW_B - C_a - 2C_b \tag{8}$$

Equation (8) shows us that when there is no chance for settlement the threat value is dependent on the probability of winning the trial and the expected value of the judgment minus a function of the costs involved. When the exchange of material information isn't effective the relative optimism is increased which results in trials. Albeit more costs involved in going for trial, the litigants are willing to sacrifice more $C_a$ and $C_b$ for a probabilistic reward of $W_B$.

Now, let's go back to equation (7) and make q = 1. When private agreement is estimated to be certain the cooperative bargain (let's denote $B_C$) is dependent on the settlement benefit, $S_B$. Thus ;

$B_C$ = $W_B$(p×0) - $C_a$(0) + $S_B$(2-1) - $C_b$(2-1)

$$B_C = S_B - C_b \tag{9}$$

However, even though the relative optimism is reduced and the plaintiff only bases her bargain on the settlement benefit offered by the defendant there is still a cost of $C_b$ (bargaining cost).

A reasonable bargain is the threat value plus the cooperative surplus shared equally between the parties. Note that the cooperative bargain here is a function of the threat value under which cooperation is possible, meaning this bargain position is certain to induce cooperation. Non-cooperative bargain position is certain to induce non-settlement.

Cooperative surplus in this regard is the difference between noncooperative position and cooperative position .

Thus;

$$B_N - B_C$$
$$= pW_B - C_a - 2C_b - S_B + C_b , \tag{10}$$
$$= pW_B - S_B - C_a - C_b$$

$B_C$ is the payoff from the claim by the plaintiff through outright settlement, $B_N$ is the payoff from the claim by the plaintiff for no settlement. So the reasonable bargain is in between the two positions, thus the defendant needs to move up by the amount $\dfrac{B_N - B_C}{2}$, while the plaintiff needs to move down by the same amount.

Therefore, the reasonable bargain, $R_B$ can be calculated as ;

$$R_B = B_C + \frac{1}{2}[pW_B - S_B - C_a - C_b]$$

$$R_B = (S_B - C_b) + \frac{1}{2}[pW_B - S_B - C_a - C_b]$$

$$R_B = \frac{1}{2}[pW_B + S_B] - \frac{1}{2}[C_a + 3C_b] \tag{11}$$

It is seen from equation (11) that the reasonable bargain amount is the average of the expected judgment from the trial and the payoff from settlement minus a function of the transaction costs. This shows that the higher the transaction costs the lower the reasonable bargain, hence higher transaction costs reduce the amount of the reasonable bargain thereby inducing private agreement ,

as opposed to Coase's no transaction cost theorem. A possible explanation is that when parties in a legal dispute anticipate the ex ante costs *vis a vis* the ex post benefit , the may rather choose the option which is more present and probable with lower benefit than to take the risk of bearing higher costs in hopes of gaining a probabilistic higher benefit. Risk lovers nevertheless will like to undertake a venturesome undertaking with a higher reward (further research is needed in behavioral economics to address the possible explanations in this issue).

Too much transaction costs on the other hand may deter the litigants from engaging in negotiations. This paper asserts that an optimum transaction cost needs to be attained in order to encourage cooperation. Section 1.2 demonstrates how optimal transaction cost is more efficient in inducing cooperation in a legal process.

**Section 1.2 : Optimal Transaction Cost**

From equation (11) ,

$$\text{when the expression } \frac{1}{2}[C_b + 3C_b] = 0 \text{ ,}$$

$$R_B = \frac{1}{2}[pW_B + S_B]$$

This implies the $R_B$ is maximum under zero transaction cost and is totally dependent on the expected judgment and settlement benefit. Meaning, the absence of or very low transaction cost makes it easy for litigants to engage in several or prolonged legal disputes without having to worry about any costs. Thus inefficient in economic terms.

$$\text{when the expression } \frac{1}{2}[C_a + 3C_b] \geq \frac{1}{2}[pW_B + S_B] \text{ ,}$$

$$R_B \text{ is zero or negative}$$

A reasonable litigant wouldn't engage at all in a bargain where transaction costs are so high that there isn't any room for making a reasonable bargain.

Since most trial fees and attorney costs do not change rapidly , there is less volatility in the administrative costs , we can assume $C_a$ as a constant (at least in a relatively not very long period). The probability of winning at trial doesn't correlate with $C_b$ , so we can assume $C_b$ as a constant as well (the $S_B$ however, may change with time during a legal process due to possible psychological factors not to be discussed in this paper).

$$\text{let's say} \quad P_C = \frac{1}{2}[pW_B + S_B]$$

$$\text{and} \quad L_C = \frac{1}{2}[C_b + 3C_b] \tag{12}$$

$$\therefore$$
$$R_B = P_C - L_C \tag{13}$$

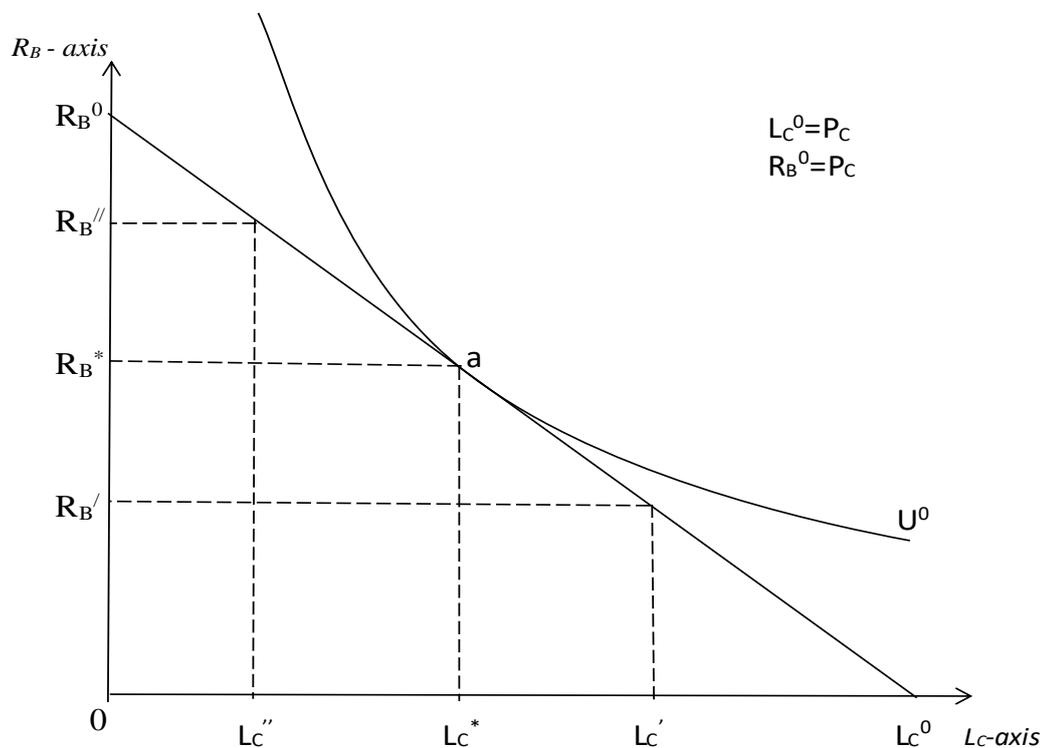

Figure (2)

As shown in figure (2) above, when $L_C = P_C$, $R_B = 0$. When $L_C = 0$, $R_B = P_C$.
If we move $L_C^0$ on the $L_C$-axis to $L_C'$ we get $R_B'$ on the $R_B$-axis. Meaning a small decrease in $L_C$ results in small increase in $R_B$. $L_C''$ corresponds with $R_B''$ which is too high and unattractive to the opponent in the negotiation. This implies very high transaction cost also causes failure in private agreements in accordance with Professor Coase's assertions. However, at point 'a' the budget line is tangent to the indifference curve; there, $L_C^*$ corresponds with $R_B^*$ showing the optimal transaction cost and the reasonable bargain of a legal negotiation.

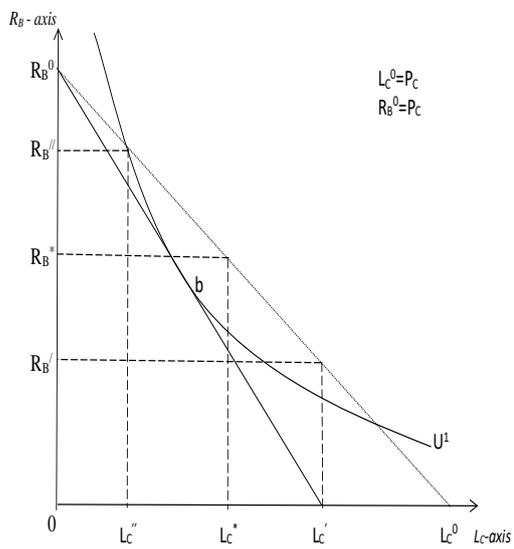

Figure (3.1)

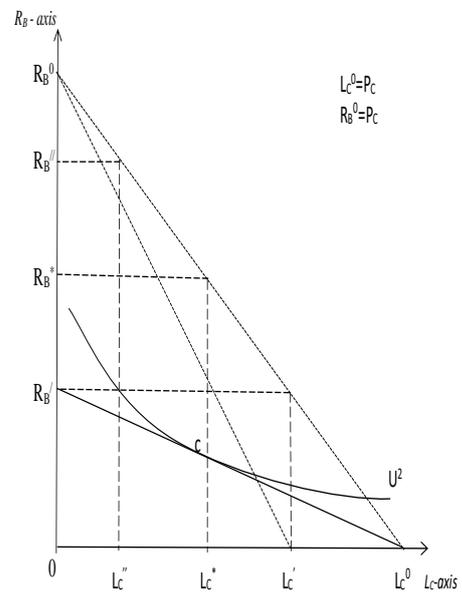

Figure (3.2)

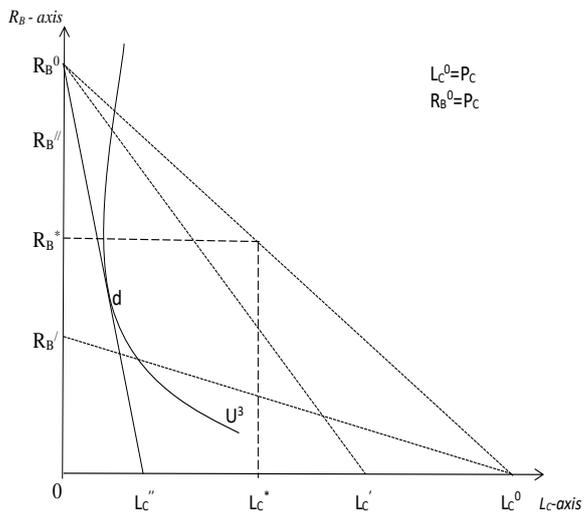

Figure (3.3)

Figure (3)

Inferring from figure (3) above ; when $R_B$ is held constant at $R_B^0$ and $L_C^0$ is reduced to $L_C^{'}$, transaction cost is reduced but expected benefit is unchanged; the utility of the new combination is $U^1$ which is clearly different from $U^0$ and obviously lower (figure 3.1). Also, when $L_C$ is held constant at $L_C^0$ and $R_B$ is moved from $R_B^0$ to $R_B^{'}$, utility changes to $U^2$ which is also different from and lower than $U^0$ (figure 3.2). Furthermore, when $R_B$ is held at $R_B^0$ and $L_C$ is moved drastically to $L_C^{''}$, utility is again changed to $U^3$ (figure 3.3).

The analysis shows that the optimal transaction cost is at $L_C^*$ with the highest utility, hence optimum transaction cost is more efficient than no or high transaction cost in estimating a reasonable bargain during a negotiation thereby inducing private agreements in a legal dispute.

$$\text{In addition, we see that as } L_C = \frac{1}{2}[C_b + 3C_b],$$

Any combination of $C_a$ and $C_b$ should result in the same optimal transaction cost, $L_C^*$. This paper recommends that courts should adjust administrative costs so that as $C_a$ increases $C_b$ decreases ; a lower $C_b$ (bargaining cost) will induce cooperation in accordance with Coase Theorem while higher $C_a$ prevents the litigants from going to trial. Further study is needed in determining how the courts could adjust the $C_a$ such that injurers (as in offenders) will not exercise less precaution because they think higher administrative costs will prevent victims from filing legal suits plus going to trial.

**Section 1.3 : Effect of Symmetric Information on a Legal Dispute**

Under symmetric information, the courts' decision is unpredictable by either party because the probability of the outcome of the judgment doesn't depend on the probabilistic strategies of the litigants. Litigants are therefore left with pure strategies to play, single or multiple Nash equilibria may be reached.

The coefficient of the winning benefit is deterministic under symmetric information, thence the reasonable bargain is deterministic as well. Perfect information is essential in negotiations for correcting false optimism thereby inducing private agreements.

A lower p in equation (11) causes the plaintiff to consider settlement.

$$\text{we can see that when p} = 0, \quad R_B = \frac{1}{2}[S_B] - \frac{1}{2}[C_a + 3C_b]$$

Meaning, the value of the claim to the plaintiff, $R_B$ is lower than the value to the defendant's offer, $S_B$. Cooperation is easily induced. Thus, the voluntary exchange of information causes settlement by correcting false optimism. Daniel Kahneman and Amos Tversky's 1979 research shows that when there is uncertainty about the future people may rather choose the option in the present than the chance to a bigger reward in the future due to loss aversion[9]. Robert Cooter and Daniel

---

[9] The theory states that people make decisions based on the potential value of losses and gains rather than the final outcome. Daniel Kahneman won the 2002 Nobel Memorial Prize in Economics for his work developing prospect theory. See; Daniel Kahneman & Amos Tversky,

Rubinfeld have demonstrated how the efforts of litigants affect the probability of going to trial ,also, how the strategy of one party is influenced by change in the strategy of the other party [10].

When p = 1 in equation (11) we find that $R_B$ is dependent on the average of $W_B$ and $S_B$ minus a function of the transaction cost. Even though $R_B$ may be high , it is still better than $B_N$ because $B_N$ is dependant directly on the $W_B$ . Albeit , $R_B$ through the rigorous calculations has been shown to be a reasonable bargain and can induce cooperation , p = 1 might wreck any hopes of negotiations as shown in 'Economic Analysis of Legal Disputes and Their Resolution' by Cooter et al[11].

**Section 1.4 : Effect of Time on Negotiations in a Legal Dispute**
Parties involved in a legal dispute stand a chance of losing economically when the legal process takes too much time. Time spent on negotiations, discovery and trials can be used for other productive purposes. Opportunity cost increases with respect to time spent in litigant.
Psychological issues, family , job , lack of finances and many others may impede the legal process or may frustrate the relative optimisms of the litigants. This paper posits that the tedious and costly nature of the legal process encourages earlier settlement and thus reduces the time used in litigation. When litigants weigh the time-value of their opportunity costs they may either reduce their bargain positions and opt for settlement or continue with the dispute if opportunity cost is less than the expected benefit from the dispute (further research is needed in determining quantitatively the effect of time-valued opportunity cost on the threat values and reasonable bargains of the parties in a legal dispute) .

## Section 2 : Axiological Evaluation of the Fair Bargain in The Legal Process
In the previous section it's been shown that under asymmetric information , the relative optimisms of litigants cause a wreck in negotiations hence the probability of settlement is reduced. The exchange of information corrects false optimism thereby inducing cooperation. Optimal transaction cost is more efficient in encouraging cooperation than no or high transaction cost. Administrative costs can be adjusted in combination with bargaining costs to formulate an optimal transaction cost which can influence the bargain positions of the litigants to encourage private agreements.
This section seeks to devise a quantitative methodology in analyzing the factors that determine the axiological threshold in the bargain position of litigants. As the previous section deduced a method of estimating the reasonable bargain under both symmetric and asymmetric information, this section uses a probabilistic approach in estimating the 'fair bargain' under uncertainty of the outcome of judgment; a method formulated from stochastic options pricing induction.

---

Prospect Theory: An Analysis of Decision Under Risk, 47 ECONOMETRICA 263 (1979)
[10] Cooter, Robert, & Daniel Rubinfeld, Economic Analysis of Legal Disputes and Their Resolution, 27 J. ECON. LIT. 1067 (1989)
[11] This ideology is the overview of the context in the same paper by Cooter et al in the footnote 10 above.

**Section 2.1 : Describing the Legal Process as an Aleatory Process in Game Theory**

Looking at figure (4) below ; there is a bargain at the decision node D which may result in non-settlement or settlement. A settlement with probability q and payoff $S_B$ to the plaintiff, non-settlement results in either trial, $T_0$ with probability 1-q or negotiation at node $P_0$ with probability q . If there is a trial, there is probability of p winning with payoff $W_B$ and 1-p losing with payoff zero. The loser might go for an appeal which doesn't have any probabilistic effect on the initial trial. At the node $P_0$ the parties engage in renegotiation where there is a probability of q to go into another renegotiation or 1-q to go to trial, $T_1$ . This palaverous conundrum could go on and on until node $P_n$ . Interactions at $P_n$ could result in another trial $T_n$ or another renegotiation $P_{n+1}$ .

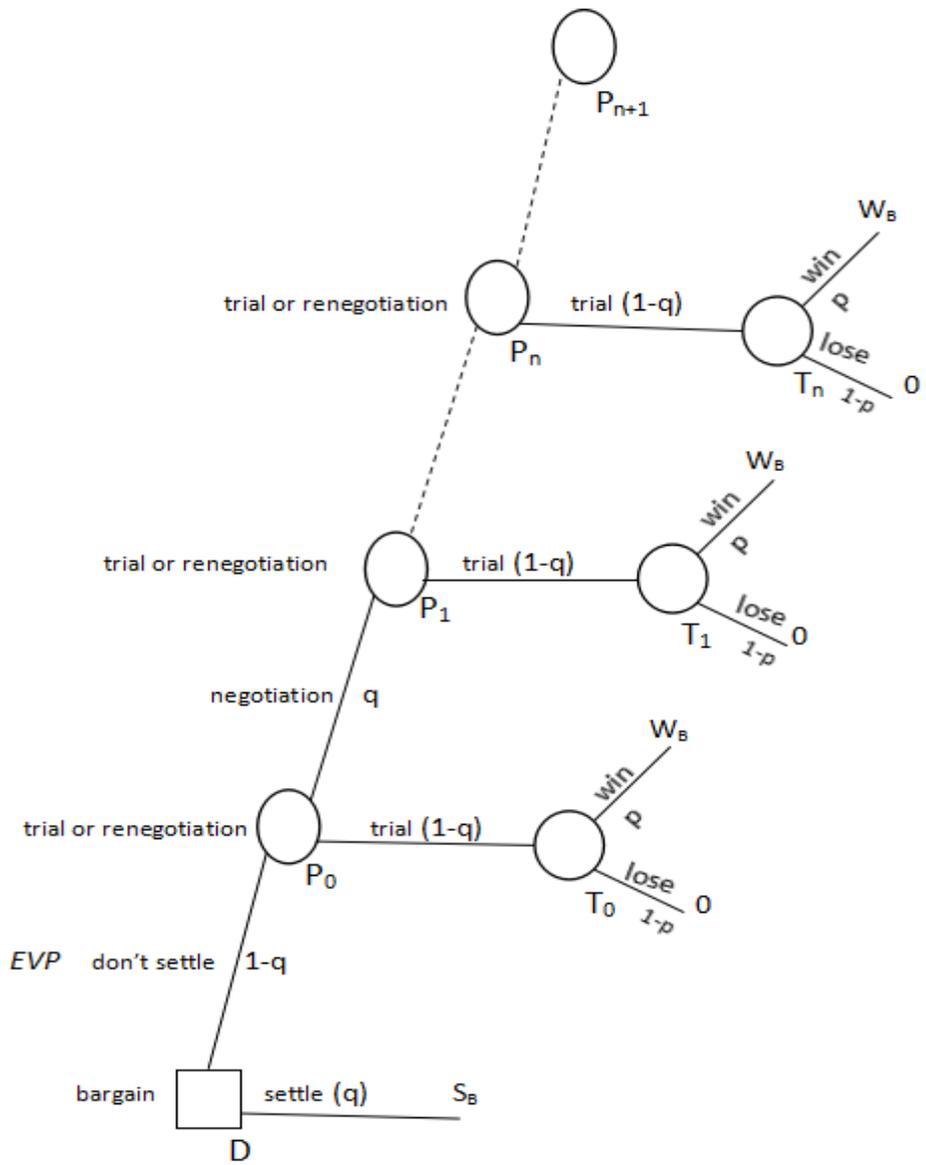

Figure (4)

The mathematical theory of Brownian Motion has been used in analyzing the haphazard movement of particles in a fluid and has also been applied to stock market financial asset pricing, options pricing and so forth[12]. We can infer from figure (4) that the probabilistic outcomes of the negotiations and trial process exhibit characteristics of Brownian motion in the sense that ;

1) the direction of the outcomes at the P nodes shows a stochastic process, just like a fair coin, under the assumption that the legal process is not biased .
2) the aleatory nature of the legal process represents a function which maps the outcomes of an unpredictable process, although some legal cases do not have numerical variables we can estimate their axiological appraisals in economic analysis . Also, the trajectories as similar as shown in figure (4) .
3) the probability of the outcomes at each stage is the same as the previous stages, if the legal process is not biased .

A stochastic process is said to follow a Geometric Brownian Motion if it satisfies the equation

$$dS_t = S_t(\mu dt + \sigma dB_t)$$

where μ is the percentage drift and σ is the percentage volatility.

The equation used in options pricing is $S_t = S_0 e^{\left(\mu - \frac{\sigma^2}{2}\right)t + \sigma dB_t}$

Options are contracts between two parties in which one party has the right but not the obligation to do something, usually to buy or sell some underlying asset. In the legal process a plaintiff has the right but not the obligation to pursue a legal claim. A legal claim can be economically expressed as an asset.

For an exercise price K, at date T, a plaintiff has the right stock price and sell it at $S_T$ on the market (in this case a legal dispute) . If $S_T > K$ then the plaintiff (owner of the option) will obtain a payoff of C at time T .

→ $C = (S_T - K)^+ = \max(S_T - K, 0)$

If $S_T \leq K$ the owner of a legal claim (the plaintiff) will not exercise her option and the payoff is zero. In a legal process, the main goal is to reduce loss as much as possible and if there is a gain on top of restitution the plaintiff deems it extra profit. Options' main goal is to gain as much profit as possible while hedging against risk. $S_T = K$ is meaningless to an options dealer and is the minimum requirement for a legal claim. $S_T < K$ however, is considered a loss in legal claims. We need to ask the question of how much a defendant is willing to pay for such a claim. This paper develops the Fair Bargain theory using the Black-Scholes-Merton options pricing model to analyse the legal process. The model uses stochastic analysis to estimate the value for the price of options.

---

[12] Brownian motion is named after the botanist Robert Brown, who first described the phenomenon in 1827, while looking through a microscope at pollen of the plant *Clarkia pulchella* immersed in water, and has since then been developed and applied in various areas such as economics, physics, mathematics, and many more. See; Ermogenous, Angeliki, "Brownian Motion and Its Applications In The Stock Market" (2006), Undergraduate Mathematics Day, Electronic Proceedings , Paper 15 , available online at; http://ecommons.udayton.edu/mth_epumd/15

Black-Scholes-Merton model is formulated on the following assumptions ( and are characteristic of the legal process ) :

1) European exercise terms are used such that options can only be exercised on the expiration date T . The American exercise terms allow options to be exercised at any time during the life of the option. In the legal process the time involved is hypothetically likened to the exercise time of an option. One could say that the plaintiff can accept the offer given by the defendant at any time so the legal process follows the American exercise terms. We should note that Black-Scholes works well for American options when interest rates and dividends are low. Also, Maria et al develops a model that generalizes the well-known Leland model with constant transaction cost function, nonlinear volatility in pricing[13]. American style options by the Black-Scholes equation, a method of free boundary problem in pricing.

At time $t \in [0,T]$ prior to the maturity time T, an American call option paying continuous dividend yield q > 0 leads to a free boundary problem.

- if $S_{f(t)} > S$ for $t \in [0,T]$ then $V(t,S) > (S-K)^+$

- if $S_{f(t)} \leq S$ for $t \in [0,T]$ then $V(t,S) = (S-K)^+$

Maria et al shows that from Kwok's work the free boundary problem for pricing American options consist in the formulation of a function V(t,S) and the early exercise boundary function $S_f$ in relation V in the Black-Scholes equation on a time depending domain shown as follows[14];
{(t,S),0<S<$S_{f(t)}$} and V(t,$S_{f(t)}$)=$S_{f(t)}$-E and ∂V(t,$S_{f(t)}$)=1 .

One could also argue that the bargain can only be reached at the end of the negotiation and whichever stage the litigation ends in the legal process is the exercise time of the option. Whichever way we argue, the Black-Scholes model applies since the legal process deals with relatively low rates and no dividends paid during the process .

2) The stock pays no dividends in the period of the option, this is very characteristic of the legal process. Payment is only made after a negotiation is attained.

3) Interest rates are constant and known. This paper posits that parties in a legal dispute should take into consideration the inflation rate of the general economy as the interest rate of the legal claim because with time the value of the legal claim is affected by the inflation rate. Remember, the main aim of the legal dispute is to restore the plaintiff to her original utility (through restitution, expectation damages or others depending on the situation), so the time-value of the claim needs to be taken into consideration.

---

[13] Black-Scholes formula is not generally applied to American call options because American options can be exercised before the expiration date, Maria et al attemps to develop an applicable model for the American options. See; Maria do Rosario Grossinho, Yaser Faghan Kord, Daniel Sevcovic,June 14, 2018,Pricing American Call Options by the Black-Scholes Equation with a Nonlinear Volatility Function , available online at; https://arxiv.org/abs/1707.00358

[14] Kwok, Y. K.: Mathematical Models of Financial Derivatives. Springer-Verlag, 1998

4) Market is efficient and there is no arbitrage. A legal dispute is assumed to be free and fair and outcomes cannot be predicted. The practice of taking advantage of a state of imbalance between two markets isn't applicable in the legal process. A plaintiff can only deal with the defendant for a particular dispute. Also, there are usually a few players in the market of optimism pricing in legal terms. In case of Mary Carter Agreement where some defendants settle by accepting a term of which is a loan by the settling defendant to the plaintiff, to be repaid by any monies recovered from the remaining defendant(s), all parties are obliged to disclose Mary Carter Agreements immediately and the non-disclosure constitute abuse of the legal process. The legal system is therefore assumed to be efficient in the following analysis.

5) No commissions are charged. It has been shown earlier in this paper that the reasonable bargain $R_B$ has already accounted for transaction costs, so if used in the Black-Scholes model there wouldn't be a need to include the transaction costs; thus a function of the administrative costs (court fees, trials, legal fees and many others) and bargaining costs (negotiation costs, settlement costs and many others). Since the transaction costs have been included in the bargain no commissions are charged in the subsequent calculations.

6) Returns on a portfolio are normally distributed for most assets that offer options.

Black-Scholes-Merton model for options pricing is stated as [15];

$$C = N(d_1)S_T - K(e^{-rT})N(d_2)$$

$$d_1 = \frac{\ln\left(\frac{S_T}{K}\right) + \left(r + \frac{\sigma^2}{2}\right)T}{\sigma\sqrt{T}}$$

$$d_2 = d_1 - \sigma\sqrt{T}$$

C = call option, S = underlying stick price, K = strike price, r = risk-free interest rate, T = time to maturity, N is a cumulative standard normal distribution

$$P[Z<x] \text{ for } Z \sim N(0,1) = \int_{-\infty}^{x} \frac{1}{\sqrt{2\pi}} e^{-\frac{u^2}{2}} du$$

In the legal process, from figures (1) and (4) and equation (11) the strike price of the claim is the reasonable bargain indicated $R_B$. The underlying stock price is the expected value of payoff from the litigation, say EVP.

---

[15] See this paper for more information about the concept and formularization of the Black-Scholes equation also known as the Black-Scholes-Merton model; Fischer Black and Myron Scholes, "The Pricing of Options and Corporate Liabilities," Journal of Political Economy 81, no. 3 (May - Jun., 1973): 637-654, available online at; https://doi.org/10.1086/260062

This implies :

$$EVP = qS_B + (1-q)E(P_0) \quad (14)$$

$$\begin{aligned}E(P_0) &= q[E(P_1)] + (1-q)[pW_B + 0] \\ &= q[E(P_1)] + (1-q)[pW_B]\end{aligned} \quad (15)$$

$$E(P_1) = q[E(P_2)] + (1-q)[pW_B] \quad (16)$$

The pattern goes on and on until $E(P_n)$, thus

$$E(P_n) = q[E(P_{n+1})] + (1-q)[pW_B] \quad (17)$$

From the Newton-Raphson method[16];

$$P_{n+1} = P_n - \left[\frac{E(P_n)}{E'(P_n)}\right]$$

A function E of the $P_{n+1}$ yields;

$$E(P_{n+1}) = E\left[P_n - \left[\frac{E(P_n)}{E'(P_n)}\right]\right]$$

Cauchy's functional equation[17] results in;

$$E(P_{n+1}) = E(P_n) - E\left[\frac{E(P_n)}{E'(P_n)}\right] \quad (18)$$

Divide equation (18) by $E(P_n)$ we get;

$$\frac{E(P_{n+1})}{E(P_n)} = \frac{E(P_n)}{E(P_n)} - \frac{E\left[\frac{E(P_n)}{E'(P_n)}\right]}{E(P_n)}$$

$$\frac{E(P_{n+1})}{E(P_n)} = 1 - \frac{E\left[\frac{E(P_n)}{E'(P_n)}\right]}{E(P_n)} \quad (19)$$

Divide equation (17) by $E(P_n)$ ;

$$\frac{E(P_n)}{E(P_n)} = q\frac{[E(P_{n+1})]}{E(P_n)} + \frac{(1-q)[pW_B]}{E(P_n)}$$

---

[16] $x_{n+1} = x_n - \frac{f(x_n)}{f'(x_n)}, n = 0,1,2,....$

see; Ben-Israel, Adi. (1965). A Newton-Raphson method for the solution of systems of equations. Journal of Mathematical Analysis and Applications. 3. 94-98. 10.1007/BF02760034

[17] Hengkrawit, Charinthip & Laohakosol, Vichian & Pianskool, Sajee. (2006). Cauchy's functional equation in restricted complex domains. International Journal of Mathematics and Mathematical Sciences. 2006. 10.1155/IJMMS/2006/69368

$$1 = q\frac{[E(P_{n+1})]}{E(P_n)} + \frac{(1-q)[pW_B]}{E(P_n)} \qquad (20)$$

Substitute equation (19) into (20) we get;

$$1 = q\left[1 - \frac{E\left[\frac{E(P_n)}{E'(P_n)}\right]}{E(P_n)}\right] + \frac{(1-q)[pW_B]}{E(P_n)}$$

$$E(P_n) = q[E(P_n)]\left[1 - \frac{E\left[\frac{E(P_n)}{E'(P_n)}\right]}{E(P_n)}\right] + (1-q)[pW_B]$$

$$E(P_n) = qE(P_n) - qE\left[\frac{E(P_n)}{E'(P_n)}\right] + (1-q)[pW_B]$$

$$E(P_n) - qE(P_n) = (1-q)[pW_B] - qE\left[\frac{E(P_n)}{E'(P_n)}\right]$$

$$E(P_n) = pW_B - \frac{qE\left[\frac{E(P_n)}{E'(P_n)}\right]}{(1-q)} \qquad (21)$$

Now, let's invert the inner variables of $E\left[\frac{E(P_n)}{E'(P_n)}\right]$, we get $E\left[\left(\frac{E'(P_n)}{E(P_n)}\right)^{-1}\right]$,

$$E\left[\frac{E(P_n)}{E'(P_n)}\right] = E\left[\left(\frac{E'(P_n)}{E(P_n)}\right)^{-1}\right]$$

From differentiation of logarithm principle, $\frac{d}{dx}\ln(f(x)) = \frac{f'(x)}{f(x)}$, we have;

$$E\left[\left(\frac{E'(P_n)}{E(P_n)}\right)^{-1}\right] = E\left[\left(\frac{d}{d(P_n)}\ln E(P_n)\right)^{-1}\right]$$

$$= E\left[\frac{d(P_n)}{d \ln E(P_n)}\right] \tag{22}$$

Substitute equation (22) into (21);

$$E(P_n) = pW_B - \frac{q}{(1-q)} E\left[\frac{P_n}{d \ln E(Pn)}\right]$$

The expression $dP_n$ which is a small change in $P_n$ is minute. That makes the expression $\frac{q}{(1-q)} E\left[\frac{dP_n}{d \ln E(Pn)}\right]$ very small comparatively. Even though change in the value of the expected payoff is small, it increases as n increases (thus as the number of negotiations and interactions increases). So we can say the expression $\frac{q}{(1-q)} E\left[\frac{dP_n}{d \ln E(Pn)}\right]$ is smallest at $P_0$, n=0. $P_0$ approaches $pW_B$ as n approaches infinity. q is also a fraction such that q≠1, if q=1 then cooperation will be assured, settlement will be induced at the node D and there wouldn't be $P_0$ at all; the game ends right there. In fact, this formula is formulated to solve disputes anticipated to go through a litigation process so q is generally small when litigation sets off, thence the product of $\frac{q}{(1-q)}$ and $E\left[\frac{dP_n}{d \ln E(Pn)}\right]$ is a very minute number.

$$\text{Therefore, } E(P_0) \approx pW_B \tag{23}$$

Substitute equation (23) into (14) we have;

$$\begin{aligned} EVP &= qS_B + (1-q)pW_B \\ &= qS_B + pW_B - qpW_B \end{aligned} \tag{24}$$

Again, we can obviously infer from equation (24) that when q = 0, EVP = $pW_B$ so a game is played, the expected value of payoff from the litigation is directly dependent on the expected payoff from trial $W_B$(winning benefit) and the probability of winning p, the litigation process sets off.

When q = 1, EVP = $S_B$, that's settlement at the node D which is same as the offer from the defendant so no bargain is needed; the WTA of the plaintiff is equal to WTP of the defendant, no game is played.

**Section 2.2 : Fair Bargain Formularization via Stochastic Options Pricing Induction**

Now, we have ;

Underlying stock price = EVP

Strike price = $R_B$

Interest rate → inflation rate = i

Time of exercise = T

Standard deviation = σ (more research is needed to develop the methodology of estimating the standard deviation in this context instrumentally)

The (options) claim Q in a legal process can be formulated in Black-Scholes formula as follows;

$$Q = N(d_1)EVP - N(d_2)R_B(e^{-iT})$$

$$d_1 = \frac{\ln\left(\frac{EVP}{R_B}\right) + \left(i + \frac{\sigma^2}{2}\right)T}{\sigma\sqrt{T}}$$

$$d_2 = d_1 - \sigma\sqrt{T}$$

$$EVP = qS_B + pW_B - qpW_B$$

$$R_B = \frac{1}{2}[pW_B + S_B] - \frac{1}{2}[C_a + 3C_b]$$

{All variables expressed in aggregate terms as opposed to values per share in financial options}

The Fair Bargain, $F_B$ in a legal dispute can therefore be estimated as
$$F_B = R_B + Q$$

This paper proposes the fair bargain as an effective bargaining position of inducing the WTA and WTP of the opponents in litigation.

Coleman et al compares the utilities of the different players in a bargain and how uncertainty affects their strategies, the paper[18] however, doesn't provide an axiological estimation of how the bargain positions (price) can induce cooperation. Cooperation is feasible from $R_B$ to $F_B$. The strength of the feasibility of cooperation is highest at $F_B$ after which cooperation becomes more and more difficult and the dispute gets closer and closer to trial.

## Section 3: Hypothesis Testing and Application

Suppose in a legal issue, the plaintiff files for a claim of $10000. The defendant offers a settlement of $5000. The probability of the plaintiff winning the case at trial is 0.6, and the probability of having a successful negotiation is 0.4. Administrative costs for trial is $1000 while costs involved in the bargaining process and settlement is $500. Inflation rate of the country of adjudication is 1.9%. It is anticipated that litigation would take 4 months to reach a straight out culmination. With an assumed standard deviation of 25%, we can estimate the fair bargain as follows;

$S_B = \$5000$, $W_B = \$10000$,
$C_a = \$1000$, $C_b = \$500$
$p = 0.6$, $q = 0.4$,
$i = 1.9\% = 0.019$, $\sigma = 25\% = 0.25$

---

[18] See; Coleman, Jules L., "A Bargaining Theory Approach to Default and Disclosure Provisions in Contract Law" (1989). Faculty Scholarship Series. Paper 4195, available online at; https://digitalcommons.law.yale.edu/fss_papers/4195/

T = 4 months = (4/12) months = 0.3333 years

$$EVP = 0.4(\$5000) + 0.6(\$10000) - 0.4 \times 0.6(\$10000)$$
$$= \$5600$$

$$R_B = \frac{1}{2}(0.6 \times \$10000 + \$5000) - \frac{1}{2}(\$1000 + 3 \times \$5000)$$
$$= \$4250$$

$$\frac{EVP}{R_B} = \frac{\$5600}{\$4250} = 1.3176 \quad ,$$

$$\ln\left(\frac{EVP}{R_B}\right) = \ln(1.3176) = 0.2758$$

$$\left(i + \frac{\sigma^2}{2}\right)T = \left(0.019 + \frac{0.25^2}{2}\right) \times 0.3333 = 0.01675$$

$$\sigma\sqrt{T} = 0.25\sqrt{0.3333} = 0.14433$$

$$d_1 = \frac{0.2758 + 0.01675}{0.14433} = 2.02695 \approx 2.03$$

$$d_2 = 2.02695 - 0.14433 = 1.88262 \approx 1.88$$

From the cumulative Normal Distribution Table[19];

$$N(d_1) = 0.9788$$
$$N(d_2) = 0.9699$$

The claim Q is given as ;

$$Q = 0.9788(\$5600) - 0.9699(\$4250)e^{(-0.019 \times 0.3333)}$$
$$= \$5481.28 - \$4250(0.99368731)(0.9699)$$
$$= \$5481.28 - \$4096.05$$
$$= \$1385.23$$

Therefore the Fair Bargain ,

$$F_B = \$4250 + \$1385.23 \approx \$5635$$

---

[19] The cumulative Normal Distribution Table shows values of d (as in $d_1$ and $d_2$) and their corresponding values of N. It can be downloaded from the University of Calgary, Department of Mathematics and Statistics Files, available online at;
https://math.ucalgary.ca/files/math/normal_cdf.pdf

Since the defendant offered $5000 as the settlement offer, $5635 is a fair bargain from the plaintiff because if negotiation fails, both parties will spend extra transactions costs of $1000 each for trial and $500 each for further negotiations. The reasonable bargain was $4250 , lower than the settlement offer, cooperation is feasible from $4250 to $5635.

Conceptual analysis of options pricing provides a wide insight into the nature of the legal process and an axiological intuition for instrumentalization of the legal process as well as procedural law.

## Conclusion

Uncertainty in the judgment, transaction costs, relative optimism, reasonable bargain affect private agreements and trials. When information is asymmetric, the relative optimisms of litigants cause a disruption in negotiations, the probability of reaching settlements is reduced, thence voluntary exchange of information not only induces private agreements but also increases efficiency in the legal process just as the undertaking of trading stocks based on non-insider information increases market efficiency.

Additionally, it is asserted in this paper that optimal transaction cost is more efficient than no or high transaction cost; and a numerical methodology is proposed on how to estimate an optimal transaction cost in a legal process. *via* backwards induction analysis this paper demonstrates the decision making process and probabilistic outcomes of the legal process in game theory and devises a method for estimating the reasonable bargain to induce the WTA(willingness to accept) of the plaintiff and WTP(willingness to pay) of the defendant in a negotiation.

Game theory mechanism per stochastic options pricing methodology is applied in the formulation of a conceptual analysis of the game played by the litigants and to estimate a 'fair' bargain in a negotiation. The above mentioned mechanism provides a wider insight into the nature of the legal process both axiologically and instrumentally.